\documentclass[runningheads]{llncs}
\usepackage{graphicx}
\usepackage{amsmath}
\usepackage{amssymb}
\usepackage{url}
\begin{document}
\title{Combining CNN and Hybrid Active Contours for Head and Neck Tumor Segmentation in CT and PET images}
%
%
\author{Jun Ma\inst{1}, Xiaoping Yang\inst{2*}}
\authorrunning{J. Ma}
%
\institute{Department of Mathematics, Nanjing University of Science and Technology\\ \email{junma@njust.edu.cn}\and
Department of Mathematics, Nanjing University\\ \email{xpyang@nju.edu.cn} (Corresponding author)
}
\maketitle              
\begin{abstract}
Automatic segmentation of head and neck tumors plays an important role in radiomics analysis. In this short paper, we propose an automatic segmentation method for head and neck tumors from PET and CT images based on the combination of convolutional neural networks (CNNs) and hybrid active contours. Specifically, we first introduce a multi-channel 3D U-Net to segment the tumor with the concatenated PET and CT images. Then, we estimate the segmentation uncertainty by model ensembles and define a segmentation quality score to select the cases with high uncertainties. Finally, we develop a hybrid active contour model to refine the high uncertainty cases.
Our method ranked second place in the MICCAI 2020 HECKTOR challenge with average Dice Similarity Coefficient, precision, and recall of 0.752, 0.838, and 0.717, respectively.
\keywords{Deep learning \and Uncertainty \and Active contours \and Characteristic function.}
\end{abstract}
\section{Introduction}
Heak and neck cancers are one of the most common cancers~\cite{CancerStatistics2020}. Extracting quantitative image bio-markers from PET and CT images has shown tremendous potential to optimize patient care, such as predicting disease characteristics \cite{lambin2012radiomics} \cite{vallieres2017radiomics}.
However, it relies on an expensive and error-prone manual annotation process of Regions of Interest (ROI) to focus the analysis. The fully automatic segmentation methods for head and neck tumors in PET and CT images are highly demanded because they will enable the validation of radiomics models on very large cohorts and with optimal reproducibility.

PET and CT modalities include complementary and synergistic information for tumor segmentation. Thus, the key is how to explore the complementary information.
Several methods have been proposed for joint PET and CT segmentation. Kumar et al. \cite{LungPET-CT} proposed a co-learning CNN to improve the fusion of complementary information in multi-modality PET-CT, which includes two modality-specific encoders, a co-learning component, and a reconstruction component.
Li et al. \cite{PET-CT-CNN-CV} proposed a deep learning based variational method for non-small cell lung cancer segmentation. Specifically, A 3D fully convolutional network (FCN) was traind on CT images to produce a probability. Then, A fuzzy variational model was then proposed to incorporate the probability map and the PET intensity image. A split Bregman algorithm was used to minimize the variational model. Recently, Andrearczyk et al. \cite{MIDL20-PET-CT} used 2D and 3D V-Net to segment head and neck tumor from PET and CT images. Results showed that using the two modalities can obtain a statistically significant improvement than using CT images or PET images only.

\begin{figure}
\centering
\includegraphics[scale=0.4]{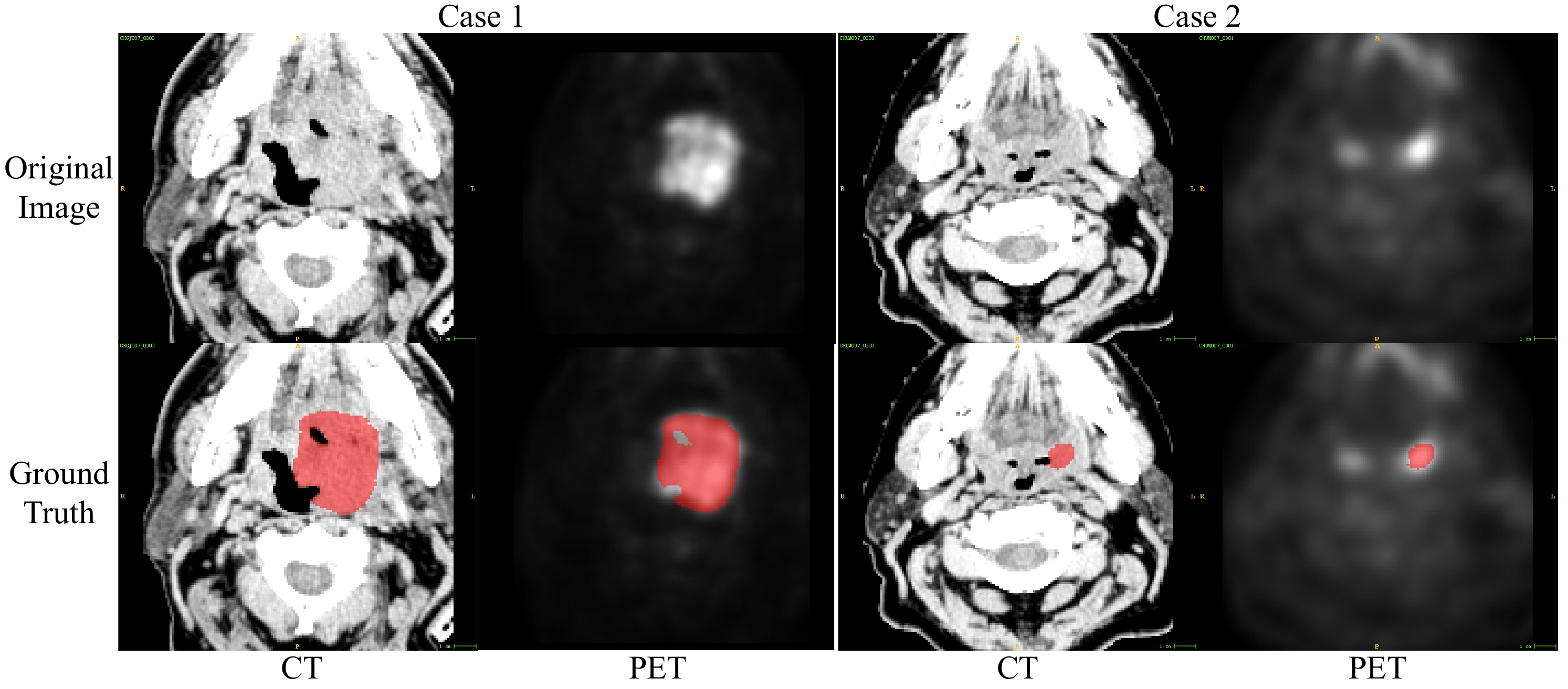}
\caption{Visual examples of PET and CT image and the corresponding ground truth.} \label{fig:eg}
\end{figure}

Active contours \cite{kass1988snakes} \cite{caselles1997GAC} \cite{chan2001CV} have been one of the widely used segmentation methods before deep learning ear.  The basic idea is to formulate the image segmentation task as an energy functional minimization problem. According to used the information in the energy functional, active contours can be classified into three categories, including edge-based active contours \cite{caselles1997GAC} that rely on image gradient information, region-based active contours \cite{li2008RSF} that rely on image-intensity region descriptors, and hybrid active contours \cite{HAC-TIP2020} \cite{zhang2008HAC} that use both image gradient and intensity information.

In this short paper, we propose an automatic segmentation method for head and neck tumors from PET and CT images based on the combination of convolutional neural networks (CNNs) and hybrid active contours. Specifically, we first introduce a multi-channel 3D U-Net to segment the tumor with the concatenated PET and CT images. Then, we estimate the segmentation uncertainty by model ensembles, and define a quality score to select the cases with high uncertainties. Finally, we develop a hybrid active contour model to refine the high uncertainty cases.


\section{Method}
\subsection{CNN backbone}
Our network backbone is the typical 3D U-Net \cite{UNet3d}.
The number of features is 32 in the first block. In each downsampling stage, the number of features is doubled. The implementation is based on nnU-Net \cite{nnunet20}.
In particular, the network input is configured with a batch size 2. The patch size is $128\times128\times128$. The optimizer is stochastic gradient descent with an initial learning rate (0.01) and a nesterov momentum (0.99). To avoid overfitting, standard data augmentation techniques are used during training, such as rotation, scaling, adding Gaussian Noise, gamma correction. The loss function is the sum between Dice loss and TopK loss~\cite{LossOdyssey}.  We train the 3D U-Net with five-fold cross validation. Each fold is trained on a TITAN V100 GPU with 1000 epochs.
The training time costs about 4 days.

\subsection{Uncertainty quantification}
We train five U-Net models with five-fold cross validation. During testing, we infer the test cases with the trained five models. Thus, each test case has five predictions. Let $p_i$ denote the predictions (Probability) of the $i-th$ model, the final segmentation $S$ can be obtained by
\begin{equation}
    S = \frac{1}{5}\sum_{i=1}^5 p_i.
\end{equation}
Then, we compute the normalized surface Dice $NSD_i$ between each prediction and the final segmentation. Details and the code are publicly available at \url{http://medicaldecathlon.com/files/Surface_distance_based_measures.ipynb}.
Finally, the uncertainty of the prediction is estimated by
\begin{equation}
    Unc = 1 - \frac{1}{5}\sum_{i=1}^5 NSD_i.
\end{equation}
If one case have a uncertainty value over 0.2, it will be selected for the next refinement.

\subsection{Refinement with Hybrid active contours}
This step aims to refine the segmentation results of the cases with high uncertainties by exploiting the complementary information among CT images, PET images, and network probabilities.
Basically, CT images can provide edge informations, and PET and network probabilities can provide location or region informations. We propose the following hybrid active contour model
\begin{equation}
    E(u) = E_{PET}(u) + E_{CT}(u) + E_{CNN}(u),
\end{equation}
where
\begin{equation}
\begin{aligned}
    E_{PET}(u; f_1, f_2) & = \int_\Omega \int_\Omega K(x,y)|I_{PET}(y)-f_1(x)|^2 u d\mathbf{x}\\
    & + \int_\Omega\int_\Omega K(x,y)|I_{PET}(y)-f_2(x)|^2 (1-u) d\mathbf{x},
\end{aligned}
\end{equation}
\begin{equation}
    E_{CT}(u) = \sqrt{\frac{\pi}{\tau}}\int_\Omega \sqrt{g_{CT}}uG_\tau*(\sqrt{g_{CT}}(1-u))d\mathbf{x}
\end{equation}
and
\begin{equation}
    E_{CNN}(u; c_1, c_2) = \int_\Omega (P_{CNN}-c_1)^2u + (P_{CNN}-c_2)^2(1-u)d\mathbf{x}.
\end{equation}
$I$ is the image intensity values, $K(x,y)$ is the Gaussian kernel function, and $c_1, c_2$ are the average image intensities inside and outside the segmentation contour, respectively.
$G_\tau$ is the Gaussian kernel, which is defined by
\begin{equation}
    G_\tau(x) = \frac{1}{4\pi\tau}\exp(-\frac{|\mathbf{x}|^2}{4\tau})
\end{equation}
The hybrid active contour model is solved by the iterative convolution-thresholding method where the details can be found at \cite{wang2017JCP,ICTM-CV,ICTM-GAC}.

\section{Experiments and results}
\subsection{Dataset}
We use the official HECKTOR dataset \cite{HECKTOR2021overview} to evaluate the proposed method.
The training data comprises 201 cases from four centers (CHGJ, CHMR, CHUM and CHUS). The test data comprise 53 cases from another center (CHUV).
Each case comprises: CT, PET and GTVt (primary Gross Tumor Volume) in NIfTI format, as well as the bounding box location and patient information. We use the official bounding box to crop all the images. We also resample the images to isotropic resolution $1mm\times1mm\times1mm$. Specifically, We use third order spline interpolation and zero order nearest interpolation for the images and labels, respectively.
Furthermore, we apply Z-score (mean subtraction and division by standard deviation) to separately normalize each PET and CT image.

\subsection{Quantitative and qualitative results}
\begin{figure}
\centering
\includegraphics[scale=0.4]{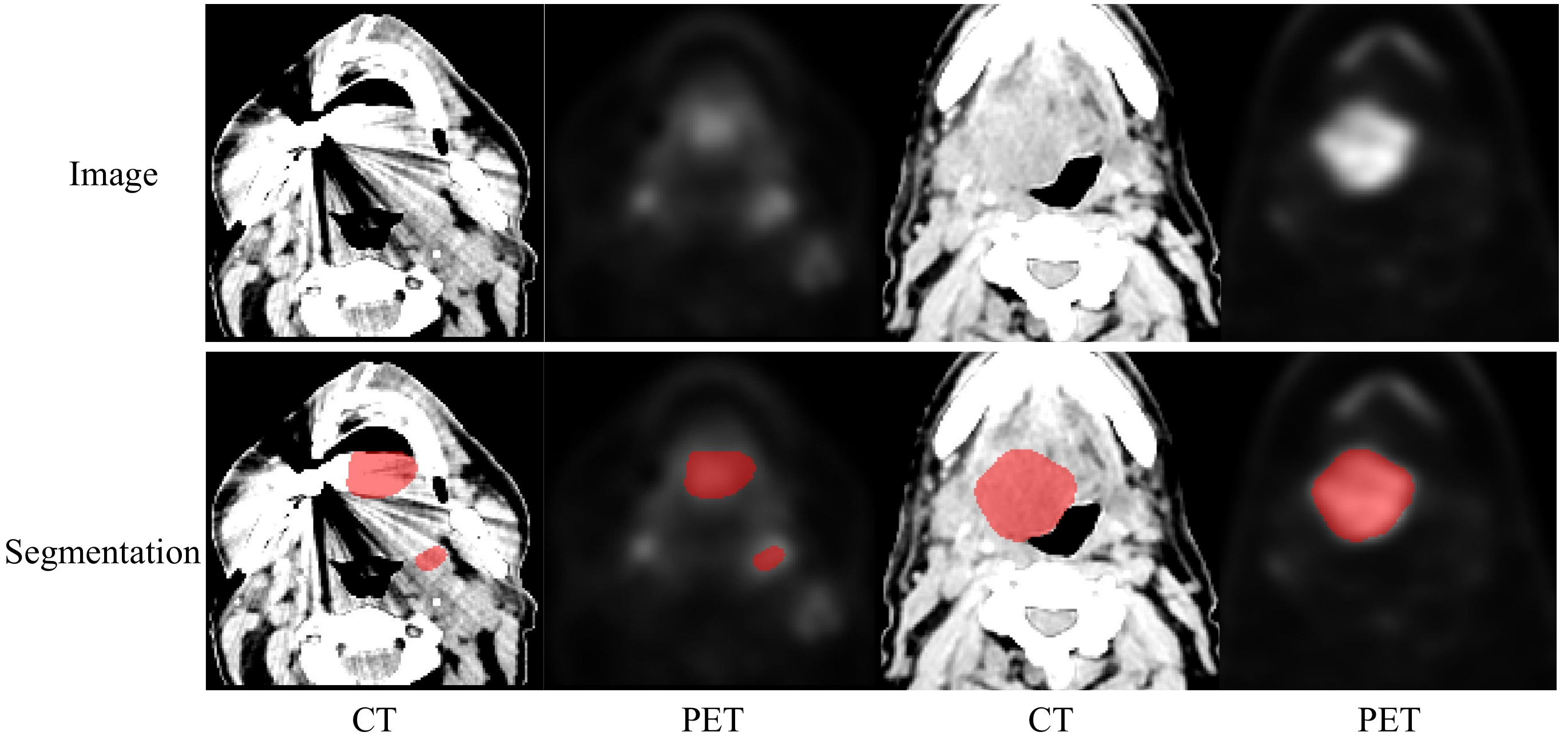}
\caption{Visual examples of segmentation results from testing set.} \label{fig:seg}
\end{figure}

Table~\ref{tab:test} and Figure \ref{fig:seg} present the quantitative and qualitative results on the testing set, respectively.
The proposed method achieved the 2nd place on the official leaderboard, which is also very close to the 1st-place performance.
The segmentation results have better precision but inferior recall, indicating that most of the segmentation results are right but some tumors are missed by the method.

\begin{table}[!h]
\caption{Quantitative results on the testing set.}\label{tab:test}
\centering
\begin{tabular}{lcccc}
\hline
Participants   & DSC & Precision & Recall & Rank  \\ \hline
andrei.iantsen & \textbf{0.759} & 0.833 & \textbf{0.740} & 1 \\
junma (\textbf{Ours})  & 0.752 & 0.838 & 0.717& 2  \\
badger         & 0.735 & 0.833 & 0.702  & 3  \\
deepX         & 0.732 & 0.785 & 0.732 & 4  \\
AIView\_sjtu   & 0.724 & \textbf{0.848} & 0.670 & 5  \\
DCPT & 0.705 & 0.765 & 0.705 & 6  \\
\hline
\end{tabular}
\end{table}

\section{Conclusion}
In this paper, we proposed a fully automatic segmentation method for head and neck tumor segmentation in CT and PET images, which combines modern deep learning methods and traditional active contours. Experiments on official HECKTOR challenge dataset demonstrate the effectiveness of the proposed method.
The main limitation of our method is the low recall, indicating that some of the lesions are missed in the segmentation results.
This would be our further work to enhance the results towards higher performance.

\section*{Acknowledgement}
This project is supported by the National Natural Science Foundation of China (No. 11531005, No. 11971229).
The authors of this paper declare that the segmentation method they implemented for participation in the HECKTOR challenge has not used any pre-trained models nor additional datasets other than those provided by the organizers.
We also thanks the HECKTOR organizers for their public dataset and hosting the great challenge.

%
%
\bibliographystyle{splncs04}
\bibliography{REF}

\end{document}